\documentstyle[glas,12pt]{article}
\newlength{\figwdth}
\input{psfig}
\begin{document}
\begin{titlepage}{GLAS-PPE/97--12}{November 1997}
\title{
The Hadronic Final States in\\
Deep Inelastic $e\gamma$ Scattering\\
at a Future e$^+$e$^-$ Linear Collider\\
}

\author{J.~Jason~Ward}

\begin{abstract}
Energy flow distributions from tagged $\gamma\gamma$ events 
generated with HERWIG are presented for $E_{beam} = 45.6$ GeV
(LEP1) and $E_{beam} = 175$ GeV (top-quark threshold at a future
e$^+$e$^-$ Linear Collider).  They have very similar shapes regardless
of the beam energy.  Using the present knowledge of the LEP 
$F_2^{\gamma}$ analyses, it is forseen that the understanding of the 
hadronic response of the forward region of the LC detector is vital 
to the development of the final state models in $\gamma\gamma$
Monte Carlos.  Such models are relied upon to extract $F_2^{\gamma}$
from the data, and hence understanding the forward region well
may be crucial in the determination of whether or not a low-$x$
rise exists in the photon structure function.
\end{abstract}
\end{titlepage}

\section{Introduction}                                                         
As the centre--of--mass energy of an e$^+$e$^-$ collider is increased, 
measurements of the photon structure function, $F_2^{\gamma}$, can be 
extended into two new kinematic regions.  A preliminary study~\cite{Red123C} 
has shown that a 500 GeV e$^+$e$^-$ linear collider with a luminosity of 
10 fb$^{-1}$ per year provides sufficient hadronic gamma-gamma events with a 
tagged electron or positron to give good statistics at both high $Q^2$ 
(10$^4$ GeV$^2$) and low Bjorken-$x$ ($0.001 < x < 0.1$).  Events at high 
$Q^2$ would allow for the QCD test of the linear rise of $F_2^{\gamma}$ 
with $\log Q^2$.  The high statistics at small $x$ presents the possibility 
of observing whether $F_2^{\gamma}$ behaves in the same way as the proton 
structure function i.e. increasing as $x$ decreases.  It is also the region 
in which theoretical predictions of $F_2^{\gamma}$ differ significantly.  
However, LEP measurements have since demonstrated that measuring $F_2^{\gamma}$ 
at low-$x$ is not trivial~\cite{Ward96,opal97-lowx} and requires a good knowledge 
of the hadronic response of the forward region detectors, in addition to having 
an electromagnetic calorimeter at small angles for tagging.  This is also true 
of the future e$^+$e$^-$ linear collider (LC) detector.  

\section{The $F_2^{\gamma}$ Low-$x$ Problem}
The measurement of the low-$x$ behaviour of $F_2^{\gamma}$ is not trivial 
for the following reason.  In the singly--tagged regime, the determination 
of $x$ requires both the $Q^2$ of the probe photon (measured from the tag)
and the invariant mass, $W$, of the hadronic final state (measured from the 
particles other than the tag).  However, the visible invariant mass is less 
than the true invariant mass mainly due to 
losses in the beam pipe and poor hadronic acceptance in the forward regions.
This results in increasing the reconstructed $x$ 
($x = Q^2/(Q^2 + W^2)$) and therefore the $x$ distribution has to be
corrected by an unfolding procedure to obtain $F_2^{\gamma}$.  
This unfolding heavily relies upon information, both before and after 
detector simulation, from the Monte Carlo that is used to model the
tagged two--photon process.  The critical point is that this Monte
Carlo must correctly model the final state, so that the particle
losses are properly accounted for.  If an unfolding Monte Carlo has 
final state particles that are more forward--going than the ones in the 
data events, the unfolding procedure can falsely increase the result
at small $x$ (and correspondingly decrease it at high $x$) and even
introduce a false low-$x$ rise into a result.  Clearly, the analysis of the 
hadronic final state is vital to the low-$x$ analysis.

The energy flow of the final state relative to the tagged electron (or 
positron) has been introduced~\cite{Ward96} and was used to demonstrate 
that the presently available tagged $\gamma\gamma$ Monte Carlos do not 
model the LEP data very well~\cite{opal97-lowx}, even in the central 
acceptance.  This results in uncertainties at low-$x$ that are too
large to be conclusive about the existence of a low-$x$ rise.

\section{The Hadronic Final State}        
\begin{figure*}[p]
\begin{center}
\vspace{-1.cm}
\mbox{\epsfysize=18.cm\epsfxsize=15.cm\epsffile
{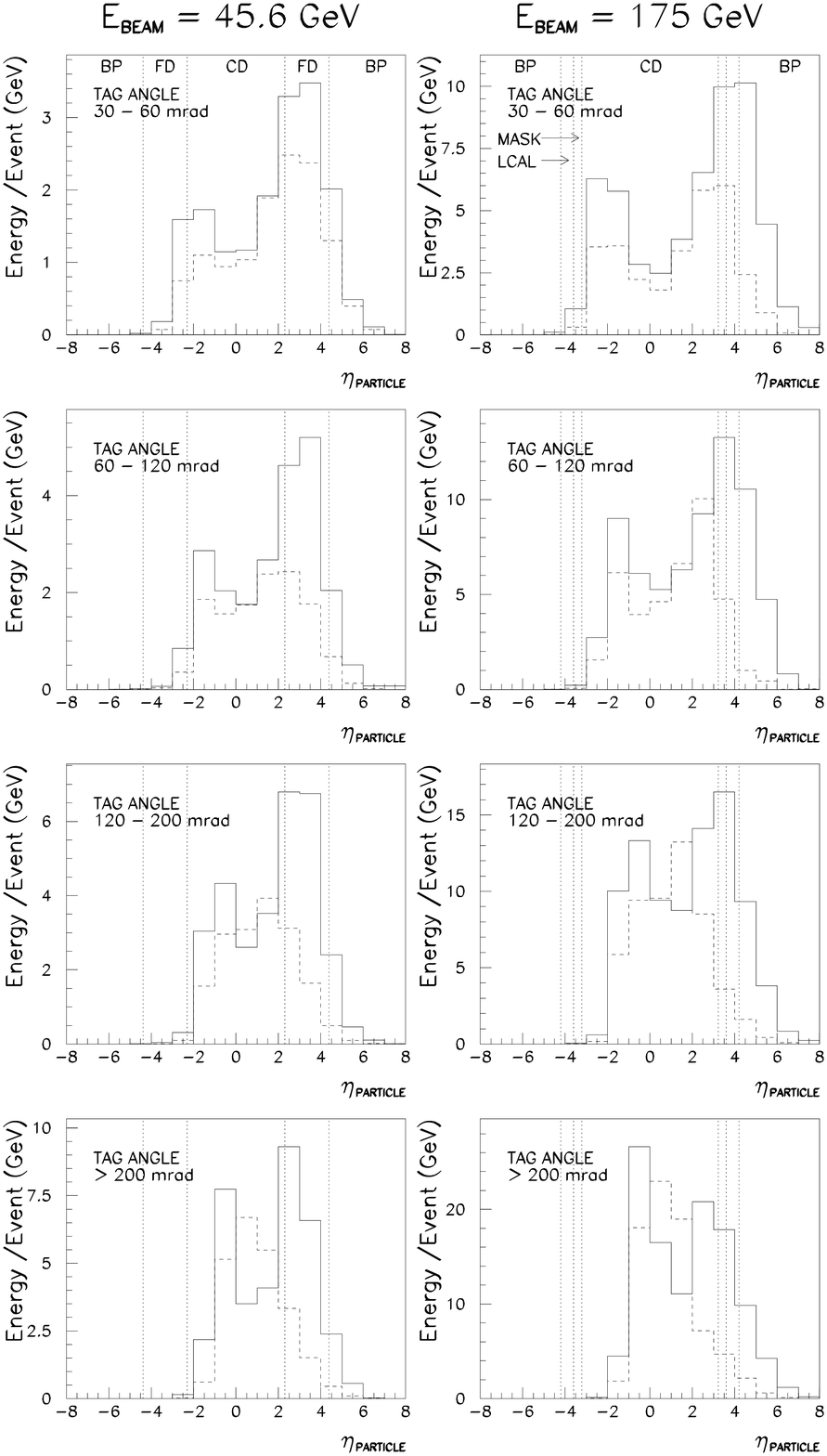}}
\end{center}
\vspace{-0.5cm}
\caption[]
{\protect\footnotesize
The $\gamma^*\gamma$ final state energy flow per event for 
$x < 0.1$ (solid line) and $x > 0.1$ (dashed line) events, as 
a function of pseudorapidity $\eta$ and the range of the tagging 
angle.  The tagged electron is always at negative pseudorapidity 
and is not shown.  The different regions in rapidity denote the 
central detector (CD), forward detector (FD) and beampipe (BP).  
For the LC detector, the arrowheads to the MASK and LCAL terminate 
in the regions of the tungsten mask and forward luminosity monitor 
inside the mask, and correspond to the TESLA configuration.  The 
pictures on the left refer to LEP1 ($E_{beam}=45.6$ GeV) and on the 
right refer to the LC ($E_{beam}=175$ GeV).
}
\label{fig:flows}
\end{figure*}

The HERWIG~\cite{Seym94} generator\footnote{Version 5.8d, IPROC=19000 (Deep 
Inelastic Scattering, all flavours, soft underlying event suppressed), 
$F_2^{\gamma} = $GRV Leading Order, Beamsstrahlung photon spectrum {\it not} 
incorporated.} has been used for this generator level study to produce
tagged $\gamma\gamma$ events 
corresponding to LEP1 and LC beam energies (45.6 and 175 GeV respectively).  
For each tagging region considered, every final state particle (other than 
the tag) has been entered into a histogram corresponding to its pseudorapidity 
($\eta = -\ln \tan (\theta/2)$), where $\theta$ is the polar angle of the 
particle measured from the direction of the beam that has produced the target 
photon) using the particle energy as a weight.  The resulting `energy flow' 
distributions are shown in Figure~\ref{fig:flows}, and are true energy 
distributions without the loss of energy due to acceptance effects.

The histograms are energy flows {\it per event} corresponding to 
$x < 0.1$ events (solid line) and $x > 0.1$ events (dashed line).
The following cuts were made:  $W_{generated} > 5$ GeV (above 
charm threshold); $E_{tag} > 0.75\ \ E_{beam}$ (N.B. changes with 
beam energy); $\theta_{tag} > 30$ mrad.  The cross--sections 
after the cuts are shown in Table~\ref{tab:xsec}.

\begin{table}[tbh]
\begin{center}
\begin{tabular}{|c|c|c|c|c|}
\hline
\hline
         & \multicolumn{4}{c|}{$\sigma_{cuts}$ (pb)} \\
\hline
         & \multicolumn{2}{c|}{$E_{beam}=45.6$ GeV} & \multicolumn{2}{c|}{$E_{beam}=175$ GeV} \\
\hline
$\theta_{tag}$ (mrad) & $x < 0.1$ & $x > 0.1$ & $x < 0.1$ & $x > 0.1$ \\ 
\hline
\hline
30--60   & 75.29 & 9.99  & 23.92 & 10.85 \\
60--120  & 13.87 & 18.39 & 3.67  & 3.71  \\
120--200 & 1.15  & 6.10  & 0.43  & 0.77  \\ 
$>200$   & 0.11  & 2.48  & 0.08  & 0.26  \\
\hline
\hline
\end{tabular}
\end{center}
\caption{\footnotesize
Cross--sections for each beam energy and tagging region 
after the cuts described in the text.  Note 
that requiring a minimum invariant mass reduces 
the cross section in the $\theta_{tag}=30-60$ mrad
range at $E_{beam}=45.6$ GeV.  
}
\label{tab:xsec}
\end{table}

The shapes of the energy flow distributions are quite similar
for the different beam energies considered.  This is an important
observation as it means that the progress and conclusions of the 
LEP studies of $F_2^{\gamma}$, especially at low-$x$, are almost 
directly applicable to the LC analysis.  It is already known 
that HERWIG generates the final state particles more forwardly 
than the LEP data~\cite{Ward96,Opal97a}, but nevertheless the 
distributions are a reasonably good approximation to those those 
expected at the LC (assuming Beamsstrahlung does not alter these
energy flow distributions too much).  

If the shapes are so similar under a change of beam energy, then 
those seen at beam energies of 45.6 GeV and 175 GeV indicate
what those at higher beam energies might be like.

There are two vital components to a low-$x$ $F_2^{\gamma}$ measurement 
at the LC.  The first is having the small angle tagger (LCAL) to 
measure the low angle tags, and hence the low-$x$ events.  This 
also maintains tagging continuity in $Q^2$ from LEP to the LC~\cite{MillVogt}.  
The second is to constrain the final state models of the $\gamma\gamma$ 
Monte Carlos with the data.  To do this, one must understand the hadronic 
response of the whole detector very well, especially the mask (if it is 
to be instrumented) and the LCAL.  This would provide vital {\it sampling} 
of the energy flow in the forward region opposite to the tag 
(approximately $3 < \eta < 4$).  One need only look at the lower 
$\theta_{tag}$ ($30 < \theta_{tag} < 120$) regions, where the 
low-$x$ cross--section is higher, to see the importance of at least sampling 
the final state at small angles, because
this is where the largest differences are seen 
between low-$x$ ($x < 0.1$) and high-$x$ ($x > 0.1$) events.
Looking at the photon structure 
problem at low-$x$ another way, one is likely to see in the final state 
the signature of a low-$x$ rise, due to the process that would be 
responsible for it, before ever unfolding to extract $F_2^{\gamma}$ 
- once again it is the knowledge of the hadronic response of the 
detector that is important.

As the tagging angle increases the situation becomes easier because 
the final state is more well contained in the central detector (the 
$p_T$ of the final state is higher in order to balance the higher 
$p_T$ of the tag).  This is reflected by the energy flow distributions
moving more towards the central detector (CD) as $\theta_{tag}$
increases.  In this tag region the measurement of $F_2^{\gamma}$ is 
therefore less model dependent, so the study of the evolution of
$F_2^{\gamma}$ at high $Q^2$ is still an achievable goal. 

\section{Conclusions}
A critical aspect of measuring $F_2^{\gamma}$ is to ensure that the 
Monte Carlos used for the unfolding procedure correctly model the 
final state.  Energy flow distributions from tagged $\gamma\gamma$ 
events have very similar shapes regardless of the beam energy in 
e$^+$e$^-$ collisions.  Using this fact, and the present knowledge 
of the LEP $F_2^{\gamma}$ analyses, it is forseen that the understanding 
of the hadronic response of the forward region of the LC detector is 
vital to the development of the final state models in $\gamma\gamma$
Monte Carlos.  This could be especially important in the determination 
of whether or not a low-$x$ rise exists in the photon structure function. 

\section{Acknowledgements}
It is a pleasure to thank David Miller of University College London, 
who supported this work and presented it on my behalf, Mike Seymour 
for his valuable help with the HERWIG generator and David Saxon for 
his helpful comments on this text.

\end{document}